\documentclass[10pt,conference]{IEEEtran}
\IEEEoverridecommandlockouts
\usepackage{cite}
\usepackage{amsmath,amssymb,amsfonts}
\usepackage{algorithmic}
\usepackage{graphicx}
\usepackage{textcomp}
\usepackage{xcolor}
\usepackage{enumitem}
\usepackage[hyphens]{url}\usepackage{hyperref}
\usepackage[text=Pre-Print,colorspec=0.9]{draftwatermark}

\def\BibTeX{{\rm B\kern-.05em{\sc i\kern-.025em b}\kern-.08em
    T\kern-.1667em\lower.7ex\hbox{E}\kern-.125emX}}
    
\newcommand{\realsearchgoal}{The goal of this research is to assist managers and other decision-makers on software projects in making informed choices about the use of different software vulnerability detection techniques through empirical analysis of the efficiency and effectiveness of each technique.}

\begin{document}

\title{Vulnerability Detection is Just the Beginning
\thanks{\copyright 2021 IEEE. Personal use of this material is permitted. Permission from IEEE must be obtained for all other uses, in any current or future media, including reprinting/republishing this material for advertising or promotional purposes, creating new collective works, for resale or redistribution to servers or lists, or reuse of any copyrighted component of this work in other works. This material is based upon work supported by the National Science Foundation under Grant No. 1909516.  Any opinions, findings, and conclusions or recommendations expressed in this material are those of the author(s) and do not necessarily reflect the views of the National Science Foundation.}
}

\author{\IEEEauthorblockN{Sarah Elder}
\IEEEauthorblockA{\textit{Department of Computer Science} \\
\textit{North Carolina State University}\\
Raleigh, USA \\
seelder@ncsu.edu}
}

\maketitle

\begin{abstract}
Vulnerability detection plays a key role in  secure software development\cite{gregoire2007secure,cruzes2017security,chatfield2017cybersecurity, morrison2018vulnerabilities}. There are many different vulnerability detection tools and techniques to choose from, and insufficient information on which vulnerability detection techniques to use and when.
\textit{\realsearchgoal}
We will examine the relationships between the vulnerability detection technique used to find a vulnerability, the type of vulnerability found, the exploitability of the vulnerability, and the effort needed to fix a vulnerability on two projects where we ensure all vulnerabilities found have been fixed. We will then examine how these relationships are seen in Open Source Software more broadly where practitioners may use different vulnerability detection techniques, or may not fix all vulnerabilities found due to resource constraints.

\end{abstract}

\begin{IEEEkeywords}
Security Management, Computer Security, Software Testing
\end{IEEEkeywords}

\section{Introduction}
Vulnerability detection plays a key role in secure software development\cite{gregoire2007secure,cruzes2017security,chatfield2017cybersecurity, morrison2018vulnerabilities}. However, there are many different vulnerability detection tools and techniques to choose from, and insufficient information on which vulnerability detection techniques to use and when.

\textit{\realsearchgoal}

The question of ``Which vulnerability detection technique should I use to reduce my security risk'' is not as straightforward as it may seem. Previous work has shown that different vulnerability detection techniques find different types of vulnerabilities\cite{austin2011onetechniquenotenough,austin2013comparison}. Hence the vulnerability detection technique that should be used will depend on the objectives of the practitioner. For example, if a practitioner has limited resources and is more concerned about remote code execution (RCE) vulnerabilities than denial of service (DoS) vulnerabilities, the practitioner may decide to focus their resources on vulnerability detection techniques that detect more RCE vulnerabilities, even if those techniques find fewer DoS vulnerabilities.

Additionally, detecting vulnerabilities does not inherently reduce risk if the vulnerabilities are not appropriately fixed or otherwise mitigated. Interviews with practitioners\cite{alomar2020you,votipka2018hackers} suggest that factors such as misalignment between security and business perspectives and miscommunication between security testers who find vulnerabilities and developers who can fix them are key to the success or failure of vulnerability detection activities. Hence we must analyze both the vulnerabilities found by vulnerability detection techniques, as well as the fixes used to mitigate those vulnerabilities and reduce security risk to understand the efficiency and effectiveness of vulnerability detection techniques.

To this end, we will examine the following research questions
\begin{itemize}
    \item RQ1: What is the efficiency of different vulnerability detection techniques?
    \item RQ2: What is the effectiveness of each vulnerability detection technique in terms of the number and type of vulnerabilities detected? 
    \item RQ3: What are the relationships between the technique with which vulnerabilities are found; the types associated with vulnerabilities; the exploitability of vulnerabilities; and the effort needed to fix vulnerabilities?
    \begin{itemize}
        \item RQ3-A: How does vulnerability exploitability relate to vulnerability type and vulnerability detection technique? 
        \item RQ3-B: How does the effort required to fix vulnerabilities, in terms of time spent performing the fix and size or complexity of the fix, relate to vulnerability type and vulnerability detection technique?
        \item RQ3-C: How does the effort required to fix vulnerabilities, in terms of time spent performing the fix and size or complexity of the fix, relate to vulnerability exploitability?
    \end{itemize}
    \item RQ4: How do vulnerabilities identified through our investigation compare with vulnerabilities taken from a larger database of open-source projects? 
\end{itemize}

The expected contributions of this work are as follows:
\begin{itemize}
    \item A decision support model, such as a decision tree, to aid practitioners in determining which vulnerability detection techniques to use based on the practitioner's goals. Additionally, the model will provide some support for prioritizing mitigation efforts based on the output of vulnerability detection tools. The decision support model will be based on our results, as well as related work and known best practices.
    \item A set of vulnerabilities from two open source projects along with the vulnerability types, the technique used to find each vulnerability, and the fix for each vulnerability.
\end{itemize}

\section{Model Parameters}
Both vulnerability risk factors such as exploitability and anticipated vulnerability fix effort contribute to vulnerability prioritization. We hypothesize that vulnerability type, exploitability, and fix effort will all vary between vulnerability detection techniques. However, we do not yet know what vulnerability types, what vulnerability exploitability, and what vulnerability fix effort will correlate with each vulnerability detection technique.  We describe the categories of vulnerability detection techniques, vulnerability types, measures for vulnerability exploitability, and measures for vulnerability fix effort below.

\subsection{Vulnerability Detection Techniques}\label{sec:param-techniques}
We characterize the vulnerability detection techniques based on the analysis they perform. Categories of analysis that we use include
\begin{itemize}[leftmargin=0.25in,after=\vspace{1pt},before=\vspace{1pt}]
\setlength{\itemindent}{-0.1in}
  \item \textbf{\emph{Static} and \emph{Dynamic} analysis}: Any sort of analysis that does not require the code to be executed may be classified as \emph{static}\cite{2013ISO29119-1}. Conversely, \emph{dynamic} analysis is any form of analysis that requires the code to be executed\cite{2013ISO29119-1}\footnote{Note, for both static analysis and dynamic analysis we are referring to the ISO/IEC/IEEE Concepts and Definitions for static testing and dynamic testing\cite{2013ISO29119-1}, respectively. For the purposes of this project, the definition for testing is generalizable to other forms of analysis such as vulnerability detection techniques.}. Analysis is either static or dynamic, not both.
  \item \textbf{\emph{Source Code} analysis}: Source code analysis is any form of analysis that requires access to source code. Analysis which does not have access to source code is sometimes referred to as ``black box'' analysis.
  \item \textbf{\emph{Tool-Based} and \emph{Manual} analysis} Many analysis techniques involve the use of an automated tool such as a fuzzer, a dynamic analysis tool. Although \emph{tool-based} techniques require manual effort to configure the tools and review the results, we reserve the term \emph{manual} for analysis that does not require automated tools.
  \item \textbf{\emph{Systematic} and \emph{Exploratory} analysis} In \emph{systematic} testing, the tester methodically develops and documents, then executes a test plan\cite{smith2011systematizing,austin2011onetechniquenotenough,smith2012effective,austin2013comparison}. In \emph{exploratory} testing, on the other hand, the tester ``spontaneously designs and executes tests based on the tester's existing relevant knowledge''\cite{2013ISO29119-1}. 
\end{itemize}
We do not anticipate examining all possible combinations of analysis, since some combinations are not widely used or not applicable. For example, \emph{Systematic} and \emph{Exploratory} analysis would only apply to \emph{manual} techniques. In an initial study of Java Applications we use four different analysis techniques for vulnerability detection: Systematic Manual Penetration Testing (SMPT), Exploratory Manual Penetration Testing (EMPT), Dynamic Application Security Testing (DAST), and Static Application Security Testing (SAST). These vulnerability detection techniques and are defined by their characteristics are shown in Table \ref{tab:VulnDetectionTechniques}.

\begin{table}[htb] \renewcommand{\arraystretch}{1.3} \caption{Vulnerability Detection Techniques from Initial Study} \label{tab:VulnDetectionTechniques}
\centering
\setlength\tabcolsep{4pt}
\begin{tabular}{ p{40pt} || p{30pt} | p{30pt} | p{40pt} |  p{40pt}}
Technique & Static / Dynamic & Source Code & Tool-Based / Manual & Systematic / Exploratory\\
\hline\hline
{SMPT} & Dynamic & No & Manual & Systematic\\\hline
{EMPT} & Dynamic & No & Manual & Exploratory\\\hline
{DAST} & Dynamic & No & Tool-Based & N/A\\\hline
{SAST} & Static & Yes & Tool-Based & N/A\\
\end{tabular}
\end{table}
\subsection{Vulnerability Types}\label{sec:param-types}
We classify vulnerabilities themselves by type using two categorizations. First, we use the vulnerability types from the Common Weakness Enumeration (CWE)\footnote{\url{https://cwe.mitre.org/}} list. According to the CWE website, CWE ``is a community-developed list of software and hardware weakness types.'' CWE is used in academic literature\cite{austin2011onetechniquenotenough,austin2013comparison,li2017large}, by government organizations such as for the U.S. National Vulnerability Database (NVD)\footnote{\url{https://nvd.nist.gov/vuln/categories}}, and in industry. All of the tools used in our research linked their alerts to CWE types. When we refer to ``vulnerability types'' we are referring to CWEs. Second, once we have identified vulnerability fixes, we determine if the vulnerability was an error of omission, e.g. input that is not sanitized at all; or an error of commission, e.g. input that is sanitized but is not sanitized correctly and can be successfully attacked.

\subsection{Vulnerability Exploitability}\label{sec:param-exploitability}
We examine vulnerability exploitability along two measures. The first exploitability measure is whether the vulnerability is ``able to be exploited'' through any means. This is a boolean measure. The second exploitability measure we will examine will assess the difficulty of exploiting a vulnerability. An example metric that could be used to assess the difficulty of exploiting a vulnerability is the Base Exploitability measure from the Common Vulnerability Scoring System (CVSS)\footnote{\url{https://www.first.org/cvss/}}. The CVSS exploitability metric is based on the context in which an attack on the vulnerability is possible, referred to as the Attack Vector (AV); and ``the conditions beyond the attacker’s control that must exist in order to exploit the vulnerability''\footnote{\url{https://www.first.org/cvss/v3.1/specification-document}}, referred to as the Attack Complexity (AC). 

\subsection{Vulnerability Fix Effort}\label{sec:param-fixes}
We  measure the effort required to fix vulnerabilities by measuring the amount of time it takes individuals to fix vulnerabilities in terms of minutes and hours. As a result of our initial study, we have already identified over 200 exploitable vulnerabilities in one SUT we have begun to fix, measuring the amount of time it takes to do so. Additionally, we will examine the size or complexity of the vulnerabilities, taking into account that size and complexity may be redundant with each other\cite{barry1981software}.


\section{Related Work}

Votipka et al\cite{votipka2018hackers} developed a model for the overall vulnerability discovery process based on interviews with both external security hackers and in-house software testers. The authors noted that although hackers and testers follow a similar process for vulnerability discovery, the different experience that hackers and testers bring to the process may explain why they achieve different results at the end of the process. The authors found that while security experience is important, a variety of experience amongst the vulnerability discovery process participants improves the overall result. Additionally, Votipka et al's interviewees asserted that compensation and motivation needed to be tailored to the organization, project, and participants. Perhaps most relevant to our own work, the authors found that the relationship between the individual who discovers a vulnerability, regardless of whether that individual is a hacker or a tester, and individual who fixes the vulnerability was considered important in ensuring that vulnerabilities were resolved appropriately.  In contrast with the work by Votipka et al, we focus on the different risks and costs of the vulnerabilities found through vulnerability detection techniques.

Several related studies have focused on one or two categories of vulnerability detection techniques, such as comparisons of DAST tools or comparisons of SAST tools. In 2010, Doup{\'e} et al.\cite{doupe2010johnny} compared several ``point and click'' DAST tools to each other. More recently, Klees et al\cite{klees2018evaluating} performed a rigorous comparison of DAST tools, providing insights on the biases and limitations of DAST tool studies. The U.S. National Institute of Standards and Technology (NIST) Software Assurance Metrics and Tool Evaluation (SAMATE) program has performed a series of Static Analysis Tool Expositions (SATE)\cite{delaitre2018sate,okun2013report,okun2011report,okun2009static}. These comparisons inform our methodology for SAST and DAST techniques, but differ from our work in that they make comparisons between tools of a similar type. Studies that compared static and dynamic analyses include a controlled experiment by Scandariato et al\cite{scandariato2013static}, which compared the use of SAST with use of DAST including a web application spider and fuzzer. Scandariato et al. performed an experiment in which nine participants performed vulnerability detection tasks. They examine  the experience of using SAST and DAST tools, and analyze the efficiency of techniques that use SAST and DAST. However, they examined efficiency and effectiveness as a function of the number of true and false positive warnings produced by these tools, without reviewing the type of vulnerabilities identified as we do in our study. Our study performs a broader comparison, looking at additional categories of vulnerability detection technique. We also examine more parameters of the vulnerabilities identified by the different vulnerability detection techniques.

An additional area of related work is the development and application of benchmarks for security testing tools, such as the 2010 work by Antunes and Viera\cite{antunes2010benchmarking} on developing a benchmark for SAST and DAST tools. As noted in the SATE V report\cite{delaitre2018sate}, benchmark studies have an important role in evaluating security testing techniques. However, the use of vulnerability detection techniques in benchmark studies may differ from how security vulnerability detection techniques would be applied in practice. Benchmarking studies assume a dataset can be created for which all vulnerabilities are known. As we found in our preliminary study, no one technique finds all types vulnerabilities - hence it is unclear how a true benchmark could be created that covers all vulnerability types. Additionally, whereas benchmarking metrics focus on vulnerability count, we examine a wider range of parameters \cite{antunes2015metrics}. 

\section{Methodology}
\subsection{Vulnerability Detection}
To begin, we will perform two studies on two different open source projects examining the efficiency and effectiveness of vulnerability detection techniques (RQ1 and RQ2). We have completed the first study on the open-source Java application OpenMRS\footnote{\url{https://openmrs.org/}}. The second study will be performed on a C/C++ application in Spring/Summer 2021.
\subsubsection{Java Application (OpenMRS)}\label{sec:method-openmrs}
In our initial study of OpenMRS, we applied SMPT, EMPT, DAST, and SAST to the SUT to produce a series of alerts, in the case of tool-based techniques; or failed test cases, in the case of manual techniques. We will refer to these alerts or failed test cases as ``failures''. The failures were then reviewed to determine the number of true positives, and to determine the number of distinct vulnerabilities indicated. For the techniques used in this study, a CWE value was assigned to each failure prior to review, and the accuracy of the CWE confirmed during the review process. Two individuals were involved in each step of the research process to improve reliability of the final dataset. 

We used data from two sources. First, we collected data from students in a graduate-level course. As part of their assignments in this course, students perform security analysis ranging from developing security requirements, to running static analysis tools, to performing exploratory testing, to fixing the vulnerabilities they find. Second, we supplemented student data with data generated by a team of three PhD students, three master's students, and one undergraduate student.

Results from our initial study are discussed in Section \ref{sec:results}. For our first study, student work was collected under NCSU Instiutional Review Board protocol 20569.  This study has been submitted to a peer-reviewed conference and is currently under review.
\subsubsection{C/C++ Application}
We have not yet determined which application will be used to expand our dataset to a C or C++ application. We anticipate using the same methodology as the previous study. We will expand the study to include additional vulnerability detection techniques. For example, we plan to expand both the Java data and the C/C++ data to include vulnerabilities identified using Interactive Application Security Testing (IAST), a tool-based technique that performs dynamic analysis but also uses source code analysis. IAST requires a user to interact with the system while the tool is running.
\subsection{Vulnerability Exploitability}\label{sec:method-exploitability}
We assume all vulnerabilities found through dynamic analysis techniques that do not have access to source code are exploitable. We will need to analyze vulnerabilities found through techniques which have access to source code such as SAST and IAST to determine if these vulnerabilities are exploitable. If this analysis is subjective, we will ensure a subset of vulnerabilities are analyzed by at least two individuals to determine if there is an acceptable level of agreement in their categorization of vulnerabilities as exploitable. Disagreements will be resolved through discussion. If agreement is low, the reviewers will review a larger subset until either the individuals have better agreement or the entire dataset has been reviewed by both individuals. The exploitability measures are discussed in Section \ref{sec:param-exploitability}. We will use this exploitability information to answer RQ3, particularly RQ3-A and RQ3-B.
\subsection{Vulnerability Fixes}
Two researchers are currently reviewing the vulnerabilities identified in Section \ref{sec:method-openmrs} to identify vulnerability fixes. While we do not have quantitative values as of January 2021, several vulnerabilities have been fixed in the latest version of the OpenMRS software which was released in April 2020. If the vulnerability has not yet been fixed, the researchers are implementing vulnerability fixes and recording the amount of time it takes to fix vulnerabilities. The code used to fix the vulnerability, whether developed by the researchers or by the original development team, will be used to compute the size or complexity of vulnerability fixes. We will continue this effort, and expand it to include vulnerabilities from the C/C++ Application. We will use this information to answer RQ3, particularly RQ3-B and RQ3-C.

\subsection{Generalizability}
Once we have completed the initial analysis for RQ1, RQ2, and RQ3, we hope to examine how the relationships identified in RQ3 are reflected in the broader open-source community. We will use data from github and other open-source software repositories. We will base our collection criteria on similar large-scale studies such as work by Li and Paxson\cite{li2017large}. While some parameters, such as the amount of time it takes an individual to fix a vulnerability, will be unavailable. Other parameters, such as the size or complexity of vulnerability fixes, will be available. We will use the available parameters to answer RQ4.

\section{Initial Results}\label{sec:results}
We have completed a preliminary study on the efficiency and effectiveness of vulnerability detection techniques in a Java application, OpenMRS. Our answers to RQ1 and RQ2 based on this initial study are as follows
\subsection{RQ1: Efficiency}
\begin{figure}[hbt]
\centering \includegraphics[width=2.2in]{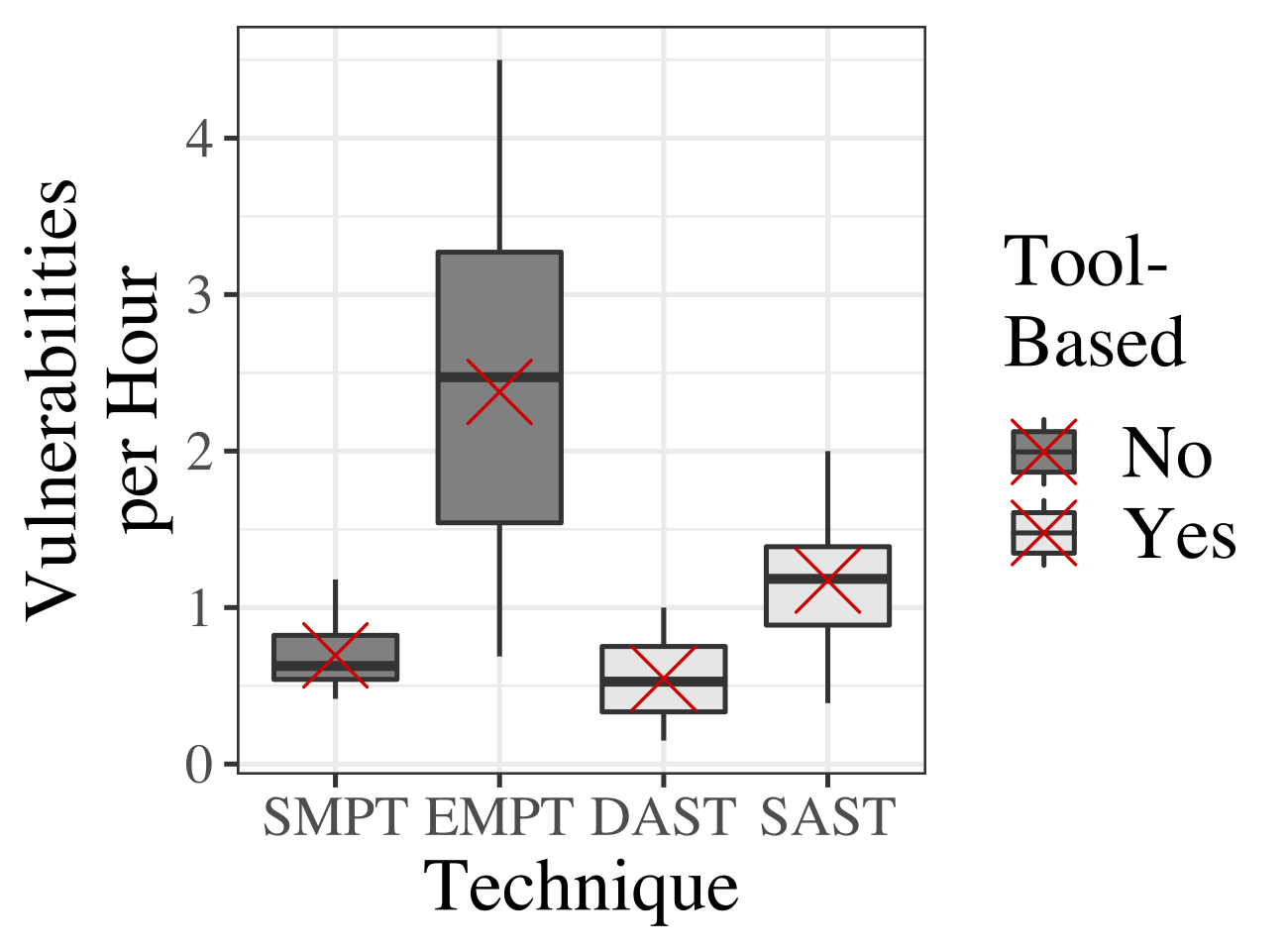} \caption{Vulnerability Detection Technique Efficiency in OpenMRS - RQ1}
\label{fig:OpenMRSEfficiency} 
\end{figure}

When efficiency is measured in terms of vulnerabilities per hour, EMPT performed notably better than the other four techniques, as can be seen in Figure \ref{fig:OpenMRSEfficiency}. Figure \ref{fig:OpenMRSEfficiency} shows the distribution of the performance of students who agreed to allow their data to be used for this research. Twelve individuals performed each technique. Several students commented that their experience with OpenMRS and better knowledge of security at the end of the course when EMPT was performed, may have contributed to their success with EMPT. This experience may help explain why EMPT also has the widest distribution in performance. Studies have found that knowledge and experience is a key factor in both functional and security exploratory testing\cite{itkonen2013role,votipka2018hackers}. Hence practitioners should consider additional factors such as the availability of professionals with experience in security or with the application itself when selecting which vulnerability detection technique to use.

\subsection{RQ2: Effectiveness}
As expected from the previous studies by Austin et al\cite{austin2011onetechniquenotenough,austin2013comparison}, all techniques found unique CWEs that were not found by the other techniques. 

\section{Timeline}
The author is a $6^{\textrm{th}}$ year PhD student at North Carolina State University (NCSU). The initial study is under revision. 

\section*{Acknowledgments}

I thank the members of the Realsearch research group for their valuable feedback on this document. This material is based upon work supported by the National Science Foundation under Grant No. 1909516.  Any opinions, findings, and conclusions or recommendations expressed in this material are those of the author(s) and do not necessarily reflect the views of the National Science Foundation. 

\bibliographystyle{IEEEtran}
\bibliography{IEEEabrv,elder}
\end{document}